\documentstyle[prl,aps,psfig]{revtex}
\begin{document}
\twocolumn[\hsize\textwidth\columnwidth\hsize\csname
@twocolumnfalse\endcsname
\title{Topological Evolution of Dynamical Networks: \\
Global Criticality from Local Dynamics$^*$}
\author{Stefan Bornholdt and Thimo Rohlf}
\address{Institut f\"ur Theoretische Physik, 
Universit\"at Kiel, Leibnizstrasse 15, D-24098 Kiel, Germany \\
{\rm (Received 6 March 2000)} 
\\
$^*$ published as Phys.\ Rev.\ Lett.\ 84 (2000) 6114. 
\\ 
}
\maketitle
\begin{abstract}
We evolve network topology of an asymmetrically connected 
threshold network by a simple local rewiring rule: 
quiet nodes grow links, active nodes lose links. 
This leads to convergence of the average connectivity 
of the network towards the critical value $K_c =2$ 
in the limit of large system size $N$. 
How this principle could generate self-organization 
in natural complex systems is discussed for two examples: 
neural networks and regulatory networks in the genome. 
\end{abstract} 
\medskip  
]

Networks of many interacting units occur in diverse areas as, 
for example, gene regulation, neural networks, 
food webs in ecology, species relationships in biological evolution, 
economic interactions, and the organization of the internet. 
For studying statistical mechanics properties of such 
complex systems, discrete dynamical networks provide a simple 
testbed for effects of globally interacting
information transfer in network structures. 

One example is the threshold network with sparse asymmetric connections.  
Networks of this kind were first studied as diluted, non-symmetric spin 
glasses \cite{D87} and diluted, asymmetric neural networks 
\cite{DGZ87,KZ87}. 
For the study of topological questions in networks, a version  
with discrete connections $c_{ij}=\pm1$ is convenient and 
will be considered here. It is a subset of Boolean networks 
\cite{K69,K90} with similar dynamical properties. 
Random realizations of these networks exhibit complex 
non-Hamiltonian dynamics including transients and limit cycles 
\cite{K88a,B96}. 
In particular, a phase transition is observed at a critical 
average connectivity $K_c$ with lengths of transients and 
attractors (limit cycles) diverging exponentially with system size for 
an average connectivity larger than $K_c$. A theoretical 
analysis is limited by the non-Hamiltonian character of 
the asymmetric interactions, such that standard tools of 
statistical mechanics do not apply \cite{D87}.  
However, combinatorial as well as numerical methods provide 
a quite detailed picture about their dynamical properties
and correspondence with Boolean Networks  
\cite{K88a,B96,DP86,DS86,DW86,DF86,K88b,F88,FK88,B98}.  

While basic dynamical properties of interaction networks 
with fixed architecture have been studied with such models, 
the origin of specific structural properties of networks 
in natural systems is often unknown. For example, the observed  
average connectivity in a nervous structure or in a biological
genome is hard to explain in a framework of networks with 
a static architecture. For the case of regulation networks 
in the genome, Kauffman postulated that gene regulatory networks
may exhibit properties of dynamical networks near criticality 
\cite{K69,K93}. However, this postulate does 
not provide a mechanism able to generate an average 
connectivity near the critical point. 
An interesting question is whether connectivity may be 
driven towards a critical point by some dynamical 
mechanism. In the following we will sketch such an approach  
in a setting of an explicit evolution of the connectivity 
of networks. 

Network models of evolving topology, in general, have been studied 
with respect to critical properties earlier in other areas, 
e.g., in models of macro-evolution \cite{Sole}. 
Network evolution with a focus on gene regulation  
has been studied first for Boolean networks in \cite{BS98}  
observing self-organization in network evolution, 
and for threshold networks in \cite{BS00}.  
Combining the evolution of Boolean networks with game theoretical 
interactions is used to model networks in economy \cite{PBC00}. 

In a recent paper Christensen et al. \cite{CDKS98} introduce 
a static network with evolving topology of undirected links 
that explicitly evolves towards 
a critical connectivity in the largest cluster of the network. 
In particular they observe for a neighborhood-oriented rewiring 
rule that the connectivity of the largest cluster evolves towards 
the critical $K_c=2$ of a marginally connected network.    
Motivated by this work we here consider the topological 
evolution of threshold networks with asymmetric links 
to study how local rules may affect global connectivity of a network, 
including the entire set of clusters of the network. 
In the remainder of this 
Letter we define a threshold network model with a local,
topology-evolving rule. Then numerical results are presented that 
indicate an evolution of topology towards a critical connectivity 
in the limit of large system size. Finally, we discuss these results 
with respect to other mechanisms of self-organization and point to  
possible links with interaction networks in natural systems. 

Let us consider a network of $N$ randomly interconnected binary 
elements  with states $\sigma_i=\pm1$. For each site $i$, its state 
at time $t+1$ is a function of the inputs it receives from other 
elements at time $t$:
\begin{eqnarray} 
\sigma_i(t+1) = \mbox{sgn}\left(f_i(t)\right) 
\end{eqnarray}  
with 
\begin{eqnarray} 
f_i(t) = \sum_{j=1}^N c_{ij}\sigma_j(t) + h.  
\end{eqnarray}
The interaction weights $c_{ij}$ 
take discrete values $c_{ij}=\pm1$, with $c_{ij} = 0$ if site 
$i$ does not receive any input from element $j$. In the following, 
the threshold parameter $h$ is set to zero. 
The dynamics of the network states is generated by iterating 
this rule starting from a random initial condition, eventually 
reaching a periodic attractor (limit cycle or fixed point). 

Then we apply the following local rewiring rule to a randomly 
selected node $i$ of the network: 
\\ {\bf 
If node $i$ does not change its state during the attractor, 
it receives a new non-zero link $c_{ij}$ from a random node $j$.  
If it changes its state at least once during the attractor, 
it loses one of its non-zero links $c_{ij}$.  
} \\   
Iterating this process leads to a self-organization of the 
average connectivity of the network. 

To be more specific, let us now describe one of several possible 
realizations of such an algorithm in detail. 
We define the average activity $A(i)$ of a site $i$ 
\begin{eqnarray} 
A(i) = \frac{1}{T_2-T_1}\sum_{t=T_1}^{T_2}\sigma_i(t) 
\end{eqnarray}
where the sum is taken over the dynamical attractor 
of the network defined by $T_1$ and $T_2$. For practical purposes, 
if the attractor is not reached after $T_{max}$ updates, $A(i)$ 
is measured over the last $T_{max}/2$ updates. This avoids 
exponential slowing down by long attractor periods for 
an average connectivity $K>2$. 
The algorithm is then defined as follows: 

(1) Choose a random network with an average connectivity $K_{ini}$. 

(2) Choose a random initial state vector 
$\vec{\sigma}(0)=$ $(\sigma_1(0),...,\sigma_N(0) )$. 

(3) Calculate the new system states $\vec{\sigma}(t),\quad t=1,...,T $ 
according to eqn.\ (2), using parallel update of the $N$ sites. 

(4) Once a previous state reappears (a dynamical attractor is reached) 
or otherwise after $T_{max}$ updates the simulation is stopped. 
Then change the topology of the network according to the following 
local rewiring rule: 

(5) A site $i$ is chosen at random and its average activity $A(i)$ 
is determined. 

(6) If $|A(i)|=1$, $i$ receives a new link $c_{ij}$ from a site $j$ 
selected at random, choosing $c_{ij}=+1$ or $-1$ with equal 
probability. If $|A(i)|<1$, one of the existing non-zero links of 
site $i$ is set to zero. 

(7) Finally, one non-zero entry of the connectivity-matrix 
is selected at random and its sign reversed.

(8) Go to step number 2 and iterate. 

\noindent 
The fluctuations introduced in step 7 as random spin reversals 
are motivated by structurally neutral noise often observed in natural 
systems. Omitting this step does not change the basic behavior of the 
algorithm, however, the distribution of the number of inputs per node 
then evolves away from a Poissonian, thereby increasing the fraction 
of nodes with many inputs. The resulting dynamics only differs from 
the original algorithm in a slightly larger connectivity $K_{ev}$ 
of the evolved networks.  
This effect vanishes $\sim 1/N$ with increasing system size.   

The typical picture arising from the model as defined above 
is shown in Fig.\ 1 for a system of size $N=1024$. 
\begin{figure}[htb]
\let\picnaturalsize=N
\def\picsize{85mm}
\def\picfilename{fig1.ps}
\ifx\nopictures Y\else{\ifx\epsfloaded Y\else\input epsf \fi
\let\epsfloaded=Y
\centerline{\ifx\picnaturalsize N\epsfxsize \picsize\fi
\epsfbox{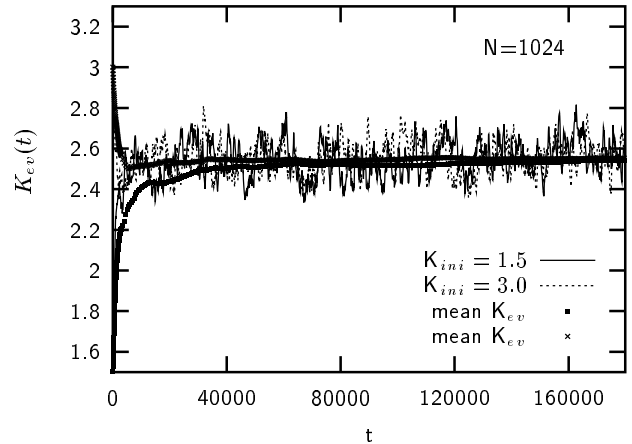}}}\fi
\caption{\small Evolution of the average connectivity of threshold
networks rewired according to the rules described in the text,
for $N=1024$ and two different initial connectivities
($K_{ini}=1.5$ and $K_{ini}=3.0$). Independent of the
initial conditions chosen at random the networks evolve
to an average connectivity $K_{ev}=2.55 \pm 0.04$.
The plot shows the time series and the corresponding
cumulative means for $K_{ev}$. The evolutionary time $t$ is discrete,
each time step representing a dynamical run on the evolved topology.
Individual runs were limited to $T_{max}=1000$ iterations. }
\end{figure}
Independent of the initial connectivity, 
the system evolves towards a statistically stationary state with 
an average connectivity $K_{ev}(N=1024)=2.55 \pm 0.04$.
With varying system size we find that with increasing $N$ the 
average connectivity converges towards $K_c$ (which, for 
threshold $h=0$ as considered here, is found at $K_c=2$), see Fig.\ 2.  
\begin{figure}[htb]
\let\picnaturalsize=N
\def\picsize{85mm}
\def\picfilename{fig2.ps}
\ifx\nopictures Y\else{\ifx\epsfloaded Y\else\input epsf \fi
\let\epsfloaded=Y
\centerline{\ifx\picnaturalsize N\epsfxsize \picsize\fi
\epsfbox{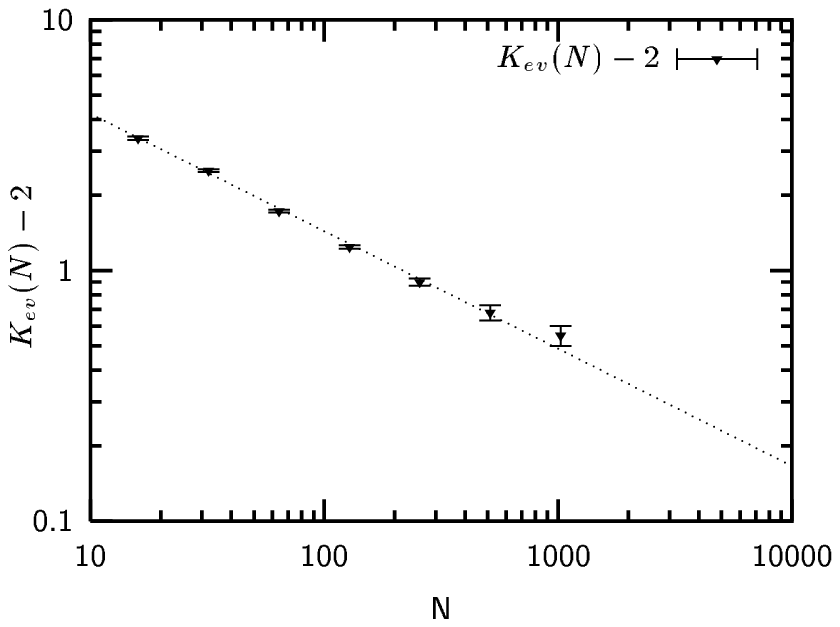}}}\fi
\caption{\small The average connectivity of the evolved networks
converges towards $K_c$ with a scaling law $\sim N^{-\delta}$,
$\delta = 0.47 \pm 0.01$. For systems with $N \le 256$ the average
was taken over $4 \cdot 10^6$ time steps, for $N=512$ and $N=1024$
over $5 \cdot 10^5$ and $2.5 \cdot 10^5$ time steps, respectively.
Finite size effects from $T_{max}=1000$ may overestimate $K_{ev}$
for the largest network shown here. }
\end{figure}
One observes the scaling relationship 
\begin{eqnarray} 
K_{ev}(N) - 2 = c\cdot N^{-\delta} 
\end{eqnarray}
with $c = 12.4 \pm 0.5$ and $\delta = 0.47 \pm 0.01$. 
Thus, in the large system size limit $N \rightarrow \infty$ 
the networks evolve towards the critical connectivity $K_c = 2$. 

The self-organization towards criticality observed in this model is 
different from currently known mechanisms exhibiting the amazingly 
general phenomenon of self-organized criticality (SOC) 
\cite{SOC,SBCW,Sole}. 
Our model introduces a (new, and interestingly different) type of 
mechanism by which a system self-organizes towards criticality, 
here $K \rightarrow K_c$. 
This class of mechanisms lifts the notions of SOC to a new level. 
In particular, it exhibits considerable robustness against noise 
in the system. 
The main mechanism here is based on 
a topological phase transition in dynamical networks. 
To see this consider the statistical properties of the average 
activity $A(i)$ of a site $i$ for a random network. 
It is closely related to the frozen 
component $C(K)$ of the network, defined as the fraction of 
nodes that do not change their state along the attractor. 
The average activity $A(i)$ of a frozen site $i$ thus obeys $|A(i)|=1$.
In the limit of large $N$, $C(K)$ undergoes a transition 
at $K_c$ vanishing for larger $K$. With respect 
to the average activity of a node, $C(K)$ equals the 
probability that a random site $i$ in the network has $|A(i)|=1$. 
Note that this is the quantity which is checked stochastically 
by the local update rule in the above algorithm. 
The frozen component $C(K,N)$ is shown for random 
networks of two different system sizes $N$ in Fig.\ 3. 
\begin{figure}[htb]
\let\picnaturalsize=N
\def\picsize{85mm}
\def\picfilename{fig3.ps}
\ifx\nopictures Y\else{\ifx\epsfloaded Y\else\input epsf \fi
\let\epsfloaded=Y
\centerline{\ifx\picnaturalsize N\epsfxsize \picsize\fi
\epsfbox{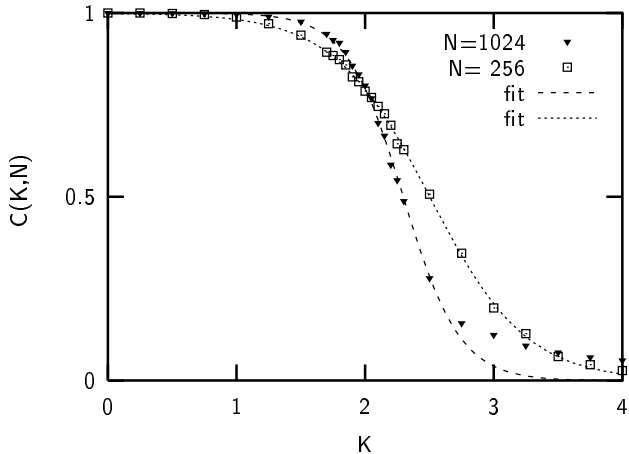}}}\fi
\caption{\small The frozen component $C(K,N)$ of random threshold
networks, as a function of the networks' average connectivities $K$.
For both system sizes shown here ($N=256$ and $N=1024$) the data were
measured along the dynamical attractor reached by the system,
averaged over 1000 random topologies for each value of $K$.
One observes a transition around a value $K = K_0$ approaching 
$K_c = 2$ for large $N$. A sigmoid function fit is also shown.
To avoid trapping in exponential divergence of attractor
periods for $K > 2$ the simulations have been limited to 
$T_{max}=10000$. The mismatch of data and fit for $N=1024$, 
$K \ge 2.75$ is due to this numerical limitation.}
\end{figure}
One finds that $C(K,N)$ can be approximated by 
\begin{eqnarray} 
C(K,N) = \frac{1}{2} \{ 1+\tanh{[-\alpha(N)\cdot(K - K_0(N)\,)]} \}.  
\end{eqnarray}
This describes the transition of $C(K,N)$ at an average connectivity 
$K_0(N)$ which depends only on the system size $N$. 
\begin{figure}[htb]
\let\picnaturalsize=N
\def\picsize{85mm}
\def\picfilename{fig4.ps}
\ifx\nopictures Y\else{\ifx\epsfloaded Y\else\input epsf \fi
\let\epsfloaded=Y
\centerline{\ifx\picnaturalsize N\epsfxsize \picsize\fi
\epsfbox{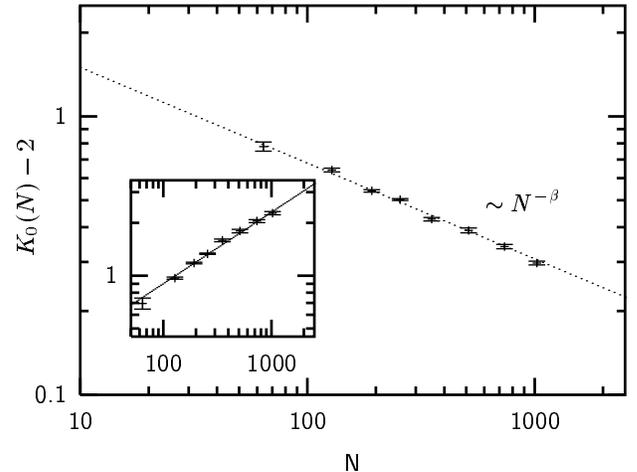}}}\fi
\caption{\small The finite size scaling of the transition value $K_0$,
obtained from sigmoidal fits as shown in FIG. 3. $K_0$ approaches
$K_c = 2$ with a scaling law $\sim N^{-\beta}$, $\beta=0.34 \pm 0.01$.
The inset shows the scaling behavior of the parameter $\alpha(N)$;
one finds $\alpha(N) \sim N^\gamma$, $\gamma = 0.41 \pm 0.014$.}
\end{figure}
One finds for the finite size scaling of $K_0(N)$ that
\begin{eqnarray} 
K_0(N) - 2 = a\cdot N^{-\beta} 
\end{eqnarray}
with $a = 3.30 \pm 0.17$ and $\beta = 0.34 \pm 0.01 $ (see Fig.\ 4), 
whereas the parameter $\alpha$ scales with system size as 
\begin{eqnarray} 
\alpha(N) = b\cdot N^\gamma 
\end{eqnarray}
with $b= 0.14 \pm 0.016$ and $\gamma = 0.41 \pm 0.01$. Thus we see 
that in the thermodynamic limit $N \rightarrow \infty$ the  
transition from the frozen to the chaotic phase becomes a sharp step 
function at $K_0(\infty) = K_c$. 
These considerations apply well to the evolving networks in 
the rewiring algorithm. 

In addition to the rewiring algorithm as described in this Letter, 
we tested a number of different versions of the model. 
Including the transient in the measurement 
of the average activity $A(i)$ results in a similar overall behavior 
(where we allowed a few time steps for the transient to decouple 
from initial conditions). Another version succeeds using 
the correlation between two sites instead of $A(i)$ as a 
mutation criterion (this rule could be called ``anti-Hebbian'' 
as in the context of neural network learning).
In addition, this version was further changed  
allowing different locations of mutated links, both, between the 
tested sites or just at one of the nodes. Some of these versions  
will be discussed in detail in a separate article \cite{RB00}.
All these different realizations exhibit the same basic behavior
as found for the model above. Thus, the mechanism proposed in this 
Letter exhibits considerable robustness. 

An interesting question is whether a comparable mechanism may 
occur in natural complex systems, in particular, whether it 
could lead to observable consequences that cannot be explained 
otherwise. 

One example where such mechanisms could occur is the 
regulation of connectivity density in neural systems. 
Activity-dependent attachment of synapses to a neuron 
is well known experimentally, for example in the form 
of the gating of synaptic changes by activity 
correlation between neurons \cite{RS79}. 
Such local attachment rules could provide a sufficient 
basis for a collective organization to occur as 
described in this Letter. 
For symmetric neural networks similar rules have been 
discussed, e.g., in the context of ``Hebbian unlearning'' 
suppressing spurious memories \cite{HFP83}. 
In the here studied asymmetric networks, however, 
such rules appear to 
generate a completely new form of self-organization dynamics.
As a consequence, an emerging average connectivity $K_{ev}$ 
could be stabilized to a specific value mostly determined by 
local properties of the dynamical elements of the system. 
It would be interesting to discuss whether synaptic density 
in biological systems could be regulated by such mechanisms.  

Another biological observable of interest is the connectivity of 
gene-gene interactions in the expression of the genome as 
first studied by Kauffman \cite{K69}. Whether this 
observable results from any such mechanism clearly is an 
open question. However, one may discuss whether biological 
evolution exerts selection pressure on the single gene level, 
that results in a selection rule similar to our algorithm; 
E.g., for a frozen regulation gene which is practically non-functional
to obtain a new function (obtain a new link), as well as for a 
quite active gene to reduce functionality (remove a link). 
First experimental estimates for the global observable of 
genome connectivity are available for E.\ coli with a value
in the range $2-3$ \cite{THPC98}.
While it is clearly too early to speculate about 
the mechanisms of global genome organization, 
it is interesting to note that the robust self-organizing 
algorithm presented here provides a mechanism that in principle 
predicts a value in this range. 

To summarize, we study topological evolution of asymmetric 
dynamical networks on the basis of a local rewiring rule.  
We observe a network evolution with the average connectivity $K$
of the network evolving towards the critical connectivity 
$K_c$ without tuning. In the limit of large system size $N$ 
this limit becomes accurate. It is well conceivable that this  
form of global evolution of a network structure towards 
criticality might be found in natural complex systems.

\end{document}